\documentclass[iop]{emulateapj}

\def\lapp{\ifmmode\stackrel{<}{_{\sim}}\else$\stackrel{<}{_{\sim}}$\fi}
\def\gapp{\ifmmode\stackrel{>}{_{\sim}}\else$\stackrel{>}{_{\sim}}$\fi}
\usepackage{multirow}
\usepackage{color}

\begin{document}

\title{{\em Swift} observations of 1FGL~J1018.6$-$5856}

\author{
Hongjun An\altaffilmark{1},
Fran{\c c}ois Dufour\altaffilmark{1},
Victoria M. Kaspi\altaffilmark{1,3},
and Fiona A. Harrison\altaffilmark{2} \\
$^1$Department of Physics, McGill University, Rutherford Physics Building, 3600 University Street,
Montreal, QC H3A 2T8, Canada\\
$^2$Cahill Center for Astronomy and Astrophysics,
California Institute of Technology, Pasadena, CA 91125, USA\\
}
\altaffiltext{3}{Lorne Trottier Chair; Canada Research Chair}

\begin{abstract}
We report on X-ray properties of the gamma-ray binary 1FGL~J1018.6$-$5856 using
observations obtained with the {\em Swift} X-ray telescope. Using 54 observations made between
MJD~55575 and 55984, we find that the X-ray flux is modulated at a period of $16.57\pm0.11$ days,
consistent with previous reports based on gamma-ray data.
We find that the X-ray maximum at phase 0 previously reported may not be a persistent feature of the source:
the dramatic increases at phase 0 were detected only for $\sim$ 100 days and not thereafter.
Rather, the persistent sinusoidal maximum seems to be at phase 0.3--0.4, and is misaligned with the gamma-ray (GeV)
peak. We also find evidence that the source's X-ray flux is correlated with the spectral hardness in the
0.5--10 keV band. Such a correlation has also been reported in the gamma-ray binaries LS~5039 and
LS~I~$+$61$^\circ$303 and can help us to understand the X-ray emission mechanisms of the sources.

\end{abstract}


\section{Introduction}
Gamma-ray binaries are a subclass of binary systems in which persistent GeV and/or TeV gamma rays are observed.
They are composed of a massive stellar companion and a compact source, and emit photons in wide range
of frequencies, from the radio to the very high energy (TeV) gamma-ray band.
Although a firm classification is still missing for some sources, we list the known
gamma-ray binaries in Table~\ref{ta:list} \citep[see][for a review]{m12}. The origin of the gamma rays is a key
puzzle in these sources.

\newcommand{\markf}{\tablenotemark{a}}
\newcommand{\markg}{\tablenotemark{b}}
\newcommand{\markh}{\tablenotemark{c}}
\newcommand{\marki}{\tablenotemark{d}}
\newcommand{\markj}{\tablenotemark{e}}
\newcommand{\markk}{\tablenotemark{f}}
\newcommand{\markl}{\tablenotemark{g}}
\begin{table*}[t]
\vspace{-0.2in}
\begin{center}
\caption{Properties of the Known Gamma-ray Binaries
\label{ta:list}}
\scriptsize{
\begin{tabular}{ccccccccc} \hline\hline
Source& Detected\markf & $P_{orb}$ & $e$\markg & Compact Source\markh & Companion & $\Gamma_{\rm X}$\marki & Corr.\markj & References \\
      & & (days) &   & & & & &  \\ \hline
1FGL~J1018.6$-$5856& R,X,G & 16.58 & $\cdots$ & Pulsar? & O6V((f)) & 1.44--1.96\markk & Yes & 1,2,3 \\
LS~5039& R,X,G,T & 3.9 & 0.35 & Pulsar? & O6.5V((f)) & 1.45--1.61 & Yes &  4,5,6,7 \\
LS~I~$+$61$^\circ$303& R,X,G,T & 26.5 & 0.55 & Pulsar? & Be & 1.7--2.0 & Yes & 8,9,10,11 \\
PSR~B1259$-$63 & R,X,G,T & $\sim$ 1240 & 0.9 & Pulsar & Be & 1.35--1.83 & No & 12,13,14,15 \\
Cyg~X-3& R,X,G,T & 0.2 & $\cdots$ & Black hole? & Wolf-Rayet & $\cdots$\markl & No & 16,17,18 \\  
HESS~J0632$+$057& R,X,T & 321 & $\cdots$ & Pulsar?  & B0pe & 1.2--1.6 & No & 19,20,21,22 \\ \hline
\end{tabular}}
\end{center}
\hspace{-0.3in}
\footnotesize{References:
[1] \citet{fermi12} 
[2] \citet{ltc+11} 
[3] \citet{hess12} 
[4] \citet{pmr+00} 
[5] \citet{tku+09} 
[6] \citet{fermi09b} 
[7] \citet{hess06} 
[8] \citet{fermi09a} 
[9] \citet{ltz+11} 
[10] \citet{LSI06} 
[11] \citet{hrl+00} 
[12] \citet{tht+11} 
[13] \citet{B1259hess05} 
[14] \citet{jml+92} 
[15] \citet{ktn+95} 
[16] \citet{wkm+94} 
[17] \citet{CygX3fermi09} 
[18] \citet{snm+11} 
[19] \citet{hsf+09} 
[20] \citet{sph+09} 
[21] \citet{bfs+11} 
[22] \citet{rt11} 
\\
See also references therein.\\}
$^{\rm a}${ Detected energy band. R=Radio, X=X-ray, G=GeV gamma ray, T=TeV gamma ray.}\\
$^{\rm b}${ Orbital eccentricity.}\\
$^{\rm c}${ Question mark if unconfirmed.}\\
$^{\rm d}${ Power-law photon index in the $\sim$ 0.5--10 keV band.}\\
$^{\rm e}${ Anti-correlation between flux and photon index in the $\sim$ 0.5--10 keV band.}\\
$^{\rm f}${ Without five flares. See text for more details.}\\
$^{\rm g}${ Continuum is not modeled with a power law.}\\
\vspace{-0.1in}
\end{table*}

There are two main models for the gamma-ray emission from these sources:
the microquasar models \citep[see][for review]{bk09} and the pulsar models \citep[see][for review]{t11}.
In the former, gamma rays are suggested to be produced in jets by Compton upscattering of the stellar
UV photons \citep[e.g.,][]{krm02, dch10}, or hadronic decay \citep[e.g.,][]{rtk+03}.
In the latter,
the gamma rays are produced by the emission from accelerated pulsar wind particles in the shock
between the pulsar and the stellar wind \citep[e.g.,][]{tak94, ta97, d06}, or
from Compton upscattering of the stellar photons by the pulsar wind particles in the pulsar wind zone
\citep[e.g.,][]{st08}. While these models give general descriptions of the gamma-ray emission
from the gamma-ray binaries, some sources show peculiar behavior
\citep[e.g., dramatic and periodic radio outbursts,
and magnetar-like bursts from LS~I~$+$61$^\circ$303,][]{hrl+00, bbc+08,dbb+08, bcd+12}
which are not presently well understood in any model \citep[e.g.,][]{mp95,tre+12}.

The 0.5--10 keV X-rays are thought to be produced via the synchrotron or the inverse Compton process
by the shock-accelerated electrons or via accretion onto the compact object.
In the wind interaction model \citep{tak94,ta97}, the X-ray flux and spectrum are expected to vary
with orbital phase. The details strongly depend on the orbital geometry and the mass loss rate of
the stellar companion \citep[see][for recent developments]{d06,chw06,bkk+08,ton+12}.
Nevertheless, \citet{ta97} suggest that the temporal behavior of the X-ray flux and
spectrum is the best diagnostic for the wind interaction models. Therefore, accurately measuring X-ray
properties of gamma-ray binaries is important to test the models and to understand physical processes
in the systems.

A significant gamma-ray and X-ray modulation from the gamma-ray binary 1FGL~J1018.6$-$5856 was
discovered by \citet{fermi12} at a period
of $16.58 \pm 0.02$ days. They noted that the orbital modulation of the X-ray and gamma-ray flux,
and the spectral variability of the gamma rays over the orbital period, are similar to those seen in
LS~5039 in general, but different in detail. The optical counterpart was spectroscopically classified as O6V((f))
using the South African Astronomical Observatory 1.9-m telescope and the 2.5-m telescope 
at the Las Campanas Observatory \citep{fermi12}, however the orbital parameters of the system are not yet well known.

Here we report on the X-ray properties of 1FGL~J1018.6$-$5856 using the {\em Swift} X-ray Telescope (XRT).
We find that the X-ray flux varies with a period
of $16.57\pm0.11$ days, consistent with the gamma-ray-measured value. We further show
evidence that the X-ray hardness is correlated with the 0.5--10 keV flux.
We compare our results with those of other gamma-ray binaries whose compact object is
(or is assumed to be) a pulsar.

\section{Observations}
\label{sec:obs}
We used 54 {\em Swift} XRT observations obtained from 2011 Jan.~14 to 2012 Feb.~27 (MJD 55575--55984),
one 20-ks {\em XMM-Newton} observation (full frame mode) in 2008 October (MJD 55066) and
one 10-ks {\em Chandra} (TE full frame mode) observation in 2010 August 17 (MJD 55425).
The 54 {\em Swift} XRT observations (all in PC mode) had different
exposures ranging from $\sim$ 0.7 ks to $\sim$ 10 ks.

We processed the {\em Swift} observations with {\ttfamily xrtpipeline} along with HEASARC
remote CALDB\footnote{http://heasarc.nasa.gov/docs/heasarc/caldb/caldb\_remote\_ac\\cess.html}
using the standard filtering procedure \citep{cps+05} to produce cleaned event files.
In each cleaned event file, we found 3 to 246 events within 20$''$ in radius centered at the source position.
The first 30 observations were analyzed and reported by \citet{fermi12}. However, we reanalyzed them for consistency.

For the {\em XMM-Newton} data, we processed the Observation Data Files (ODF) with {\ttfamily epproc} and
{\ttfamily emproc} and then applied the standard filtering procedure (e.g., flare rejection and pattern selection) of
Science Analysis System (SAS) version 11.0.0.\footnote{http://xmm.esac.esa.int/sas/}

The {\em Chandra} data were reprocessed using {\ttfamily chandra\_repro} of CIAO 4.4 along with CALDB 4.4.7
to use the most recent calibration files.
They are used for the imaging analysis only because a meaningful spectral analysis was impossible due to pile-up.

\newcommand{\markx}{\tablenotemark{a}}
\newcommand{\marky}{\tablenotemark{b}}
\newcommand{\markt}{\tablenotemark{c}}
\newcommand{\markz}{\tablenotemark{d}}
\newcommand{\markw}{\tablenotemark{e}}
\newcommand{\marku}{\tablenotemark{f}}
\newcommand{\markv}{\tablenotemark{g}}
\begin{table*}[t]
\vspace{-0.2in}
\begin{center}
\caption{Summary of spectral fit results
\label{ta:spec}}
\scriptsize{
\begin{tabular}{cccccccc} \hline\hline
Phase\markx  & Exposure & Counts\marky & $N_{\rm H}$\markt &$\Gamma$ & Flux\markz & Method & Comment \\
     & (ks)  &     & ($10^{22}\ \rm cm^{-2}) $   &     &  & \\ \hline
0.65 & 20.0 (12.1\markw) & 2528 & 0.67(5) & 1.64(7)& 1.01(3) & $\chi^2$ & {\em XMM} \\
0    & 36.9  & 768   & $\cdots$ & 1.31(8)  & 2.52(11) & $\chi^2$ & {\em Swift} combined \\
1    & 12.3  & 115   & $\cdots$ & 1.51(25) & 0.99(13) & cstat & {\em Swift} combined  \\
2    & 7.9   & 98    & $\cdots$ & 1.61(25) & 1.45(18) & cstat & {\em Swift} combined  \\
3    & 14.2  & 203   & $\cdots$ & 1.46(17) & 1.86(16) & $\chi^2$ & {\em Swift} combined  \\
4    & 8.8   & 157   & $\cdots$ & 1.44(19) & 2.01(21) & cstat & {\em Swift} combined \\
5    & 9.2   & 118   & $\cdots$ & 1.63(24) & 1.40(15) & cstat & {\em Swift} combined  \\
6    & 12.5  & 143   & $\cdots$ & 1.96(20) & 1.27(12) & cstat & {\em Swift} combined  \\
7--8 & 12.2  & 97    & $\cdots$ & 1.80(30) & 0.73(11) & cstat & {\em Swift} combined  \\
9    & 12.1  & 175   & $\cdots$ & 1.59(17) & 1.68(15) & cstat & {\em Swift} combined  \\ \hline
0\marku & 22.1 & 582 & $\cdots$ & 1.30(9)  & 3.12(16) & $\chi^2$ & {\em Swift} combined \\ 
0\markv & 14.7 & 193 & $\cdots$ & 1.50(16) & 1.49(13) & cstat & {\em Swift} combined  \\ \hline
\end{tabular}}
\end{center}
\hspace{-0.3in}
\footnotesize{Notes. Uncertainties are at the 1$\sigma$ level.\\}
$^{\rm a}${ Orbital phase as reported by \citet{fermi12}.}\\
$^{\rm b}${ Within the extraction regions and in the 0.5--10 keV. MOS1, MOS2 and PN combined for XMM.}\\
$^{\rm c}${ $N_{\rm H}$ was measured with the {\em XMM-Newton} data and fixed to $0.67\times 10^{22}\ \rm cm^{-2}$ for the {\em Swift} data fits.}\\
$^{\rm d}${ Absorption-corrected flux in the 0.5--10 keV band in units of $10^{-12}$ erg cm\textsuperscript{$-$2} s\textsuperscript{$-$1}.}\\
$^{\rm e}${ For PN detector.}\\
$^{\rm f}${ The five flares in Figures~\ref{fig:spectra}a and b.}\\
$^{\rm g}${ Without the five flares.}\\
\vspace{-0.1in}
\end{table*}

\begin{figure}
\centering
\vspace{-0.1in}
\hspace{-0.45in}
\includegraphics[width=3.6 in]{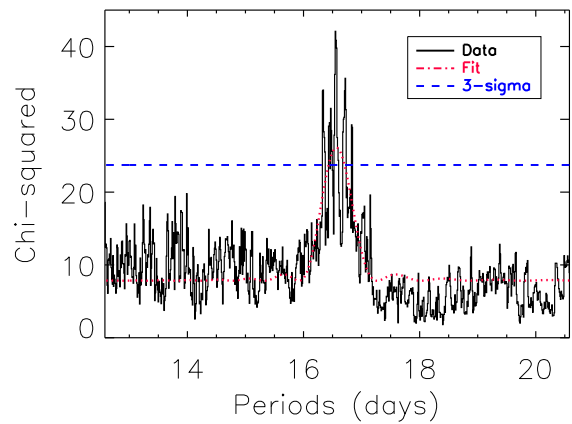} 
\vspace{-0.1in}
\figcaption{Chi-squared vs. period for the {\em Swift } X-ray data obtained using epoch-folding \citep{l87}
for a step size of 0.01 day with 8 phase bins. The fit (red) gives the best period.
\label{fig:time}
}
\end{figure}

\section{Data Analysis and Results}
\label{sec:ana}
\subsection{Imaging Analysis}
\label{imageana}
We detected the source in the 10-ks {\em Chandra} data using {\ttfamily wavdetect}
and found the source position to be R.A. = 10\textsuperscript{h}18\textsuperscript{m}55\textsuperscript{s}.62 and
Decl. = $-$58$^\circ$56$'$46$''$.06. This is consistent with what was reported by \citet{fermi12} using
{\em Swift} UVOT, the United States Naval Observatory B1.0 catalog and radio observations made with the
Australia Telescope Compact Array (ATCA): R.A. = 10\textsuperscript{h}18\textsuperscript{m}55\textsuperscript{s}.60,
Decl. = $-$58$^\circ$56$'$46$''$.2, (J2000).
We then searched for point sources that may contaminate the {\em Swift} or
{\em XMM-Newton} spectra in a circular region (radius=90$''$) and found none, which validates the extraction regions
we use below. 

We also checked if the {\em XMM-Newton}- and {\em Swift}-measured positions are consistent with the known position using
{\ttfamily edetect\_chain} and {\ttfamily wavdetect}, respectively.
The positions we found agreed with the known one within the uncertainties except for in one {\em Swift} observation,
where the source was offset by $\sim9''$ (2$\sigma$). The latter offset in a single observation is to be expected
given the large number of observations. Note that {\ttfamily wavdetect} did not detect the source in 4 {\em Swift}
observations for which the number of counts within a circle of 20$''$ radius was less than 10.
We ignored these observations for the analyses below.

\subsection{Timing Analysis}
\label{timingana}
To search for pulsations, we first applied the barycenter correction to the events
using {\ttfamily barycen} and {\ttfamily barycorr} for the {\em XMM-Newton} and the {\em Swift}
event files, respectively. We then extracted photon arrival times from the event files.

Since the source is a binary, Doppler shifting of the pulsations could broaden a periodogram made
assuming a fixed periodicity, which could reduce or entirely eliminate our sensitivity to pulses.
We consider the effects of binary orbital Doppler shifting under the assumption of a circular orbit with
30$^\circ$ inclination.
This is reasonable because the eccentricity and inclination of 1FGL~J1018.6$-$5856 are estimated to be
low under the assumption of gamma rays being produced via the inverse Compton process \citep{fermi12}.
Further, we assume the mass of the secondary star to be 25 $M_\odot$ \citep[see][for example]{pkh+96} and the
primary star to be 1.4 $M_\odot$ (for a neutron star). For the known orbital period (16.58 days),
we find that the orbital speed
of the primary star would be $\sim$ 240--250 km $\rm s^{-1}$.
In this case, the Doppler shift in the putative pulse period is
$\frac{\Delta P}{P}\sim4\times10^{-4}$
(i.e., $\sim6\times 10^{-9}$ s for a 20-ks observation for the minimum searching period of $P\sim150\ \rm ms$).
This is smaller than the independent period bins
(e.g., $P^2/T\sim1\times10^{-6}$ s for $P=150\ \rm ms$ and $T=20\ \rm ks$), and thus the blurring is not
a concern for the individual observations.
The Doppler shift over a full orbit could be as large as $\sim6\times 10^{-5}$ s (for $P=150\ \rm ms$),
hence precluding searching for pulsations by combining observations.

We searched for possible pulsations (in the 0.5--10 keV and 1--7 keV bands)
from the source using the $H$-test \citep{dsr+89}
in the individual {\em Swift} and {\em XMM-Newton} time series over the periods from the Nyquist limit
of each detector (5.2 s for MOS1 and MOS2, 146 ms for PN, 5 s for XRT) to 2000 s,
and found no significant pulsations. The most significant peak occurred at $P\simeq179\ \rm ms$
in the {\em XMM-Newton} data (0.5--10 keV),
and the probability of its occurring by chance was $\sim$ 8\%. Since this peak was not significant, we set
an upper limit on the pulsed fraction to be $F_{\rm area}\lapp49$\% or $F_{\rm rms}\lapp21$\% with 90\%
confidence, where $F_{\rm area}$ and $F_{\rm rms}$ are defined as 
$$F_{\rm area}=\frac{\sum_{i=1}^{N}(p_{i,max}-p_{i,min})}{\sum_{i=1}^{N}p_{i,min}},$$ and
$$F_{\rm rms}=\frac{\sqrt{2\sum_{k=1}^{5}((a_k^2+b_k^2)-(\sigma_{a_k}^2+\sigma_{b_k}^2))}}{a_0},$$
where $a_k=\frac{1}{N}\sum_{i=1}^{N}p_i \cos(2\pi ki/N)$, $\sigma_{a_k}$ is the uncertainty in $a_k$,
$b_k=\frac{1}{N}\sum_{i=1}^{N}p_i \sin(2\pi ki/N)$, $\sigma_{b_k}$ is the uncertainty in $b_k$,
$p_i$ is counts in $i$-th bin, and $N$ is the total number of bins
\citep[see][for more details]{gdk+10}.

\begin{figure*}
\centering
\vspace{8mm}
\begin{tabular}{cc}
\hspace{-12.00mm}
\vspace{-8.00mm}
\includegraphics[width=4. in]{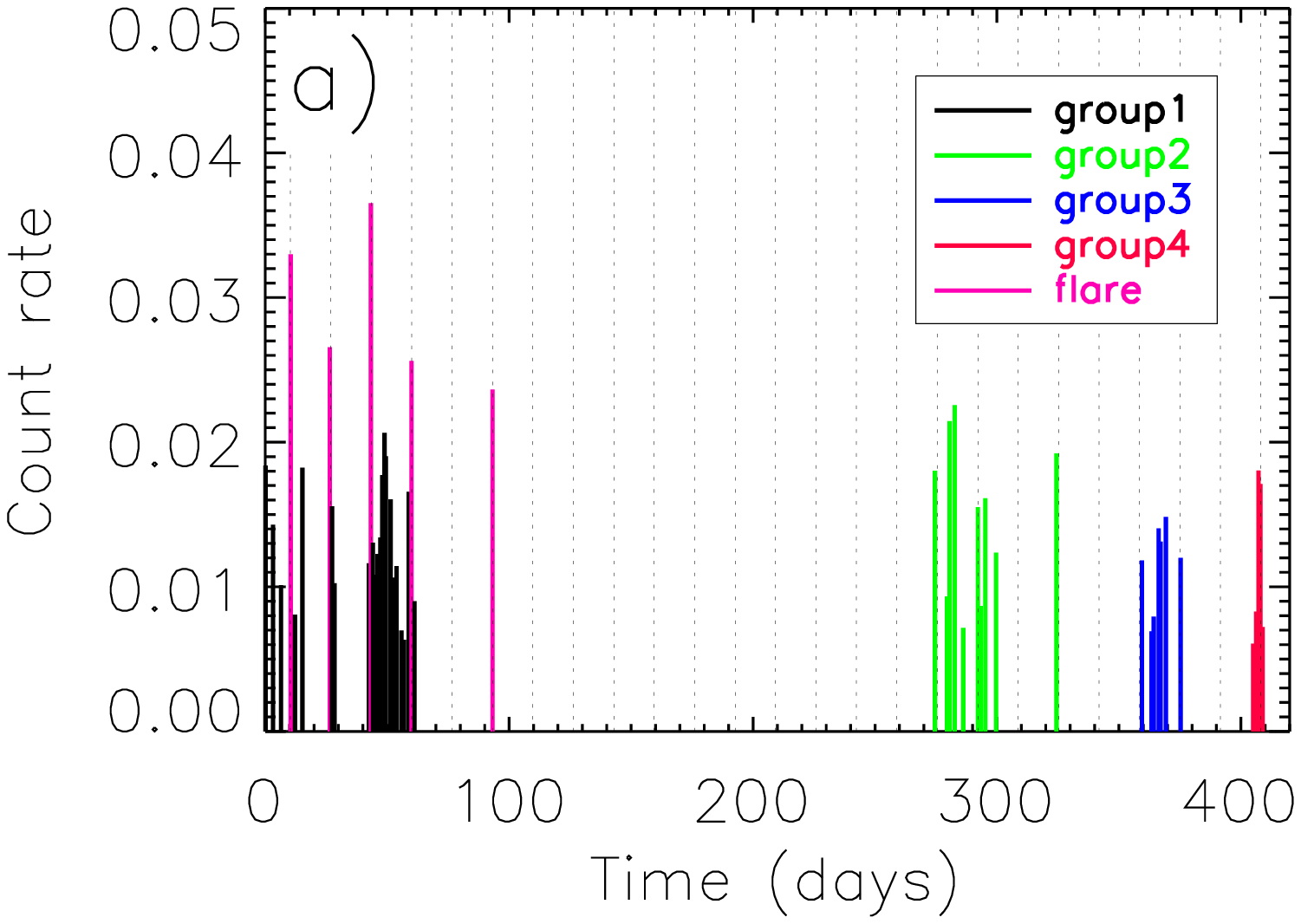} &
\hspace{-15.00mm}
\includegraphics[width=4. in]{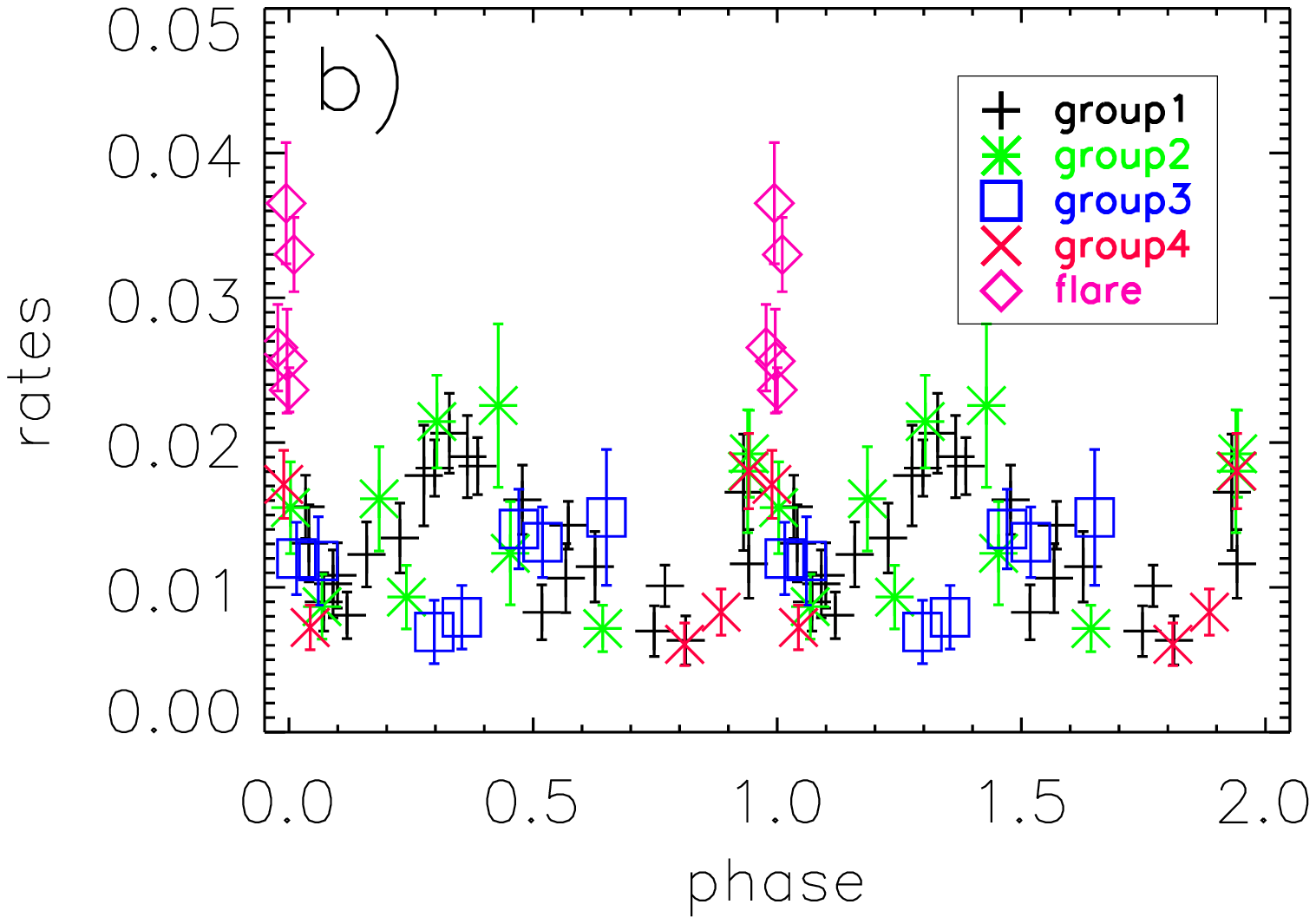} \\
\hspace{-12.00mm}
\includegraphics[width=4. in]{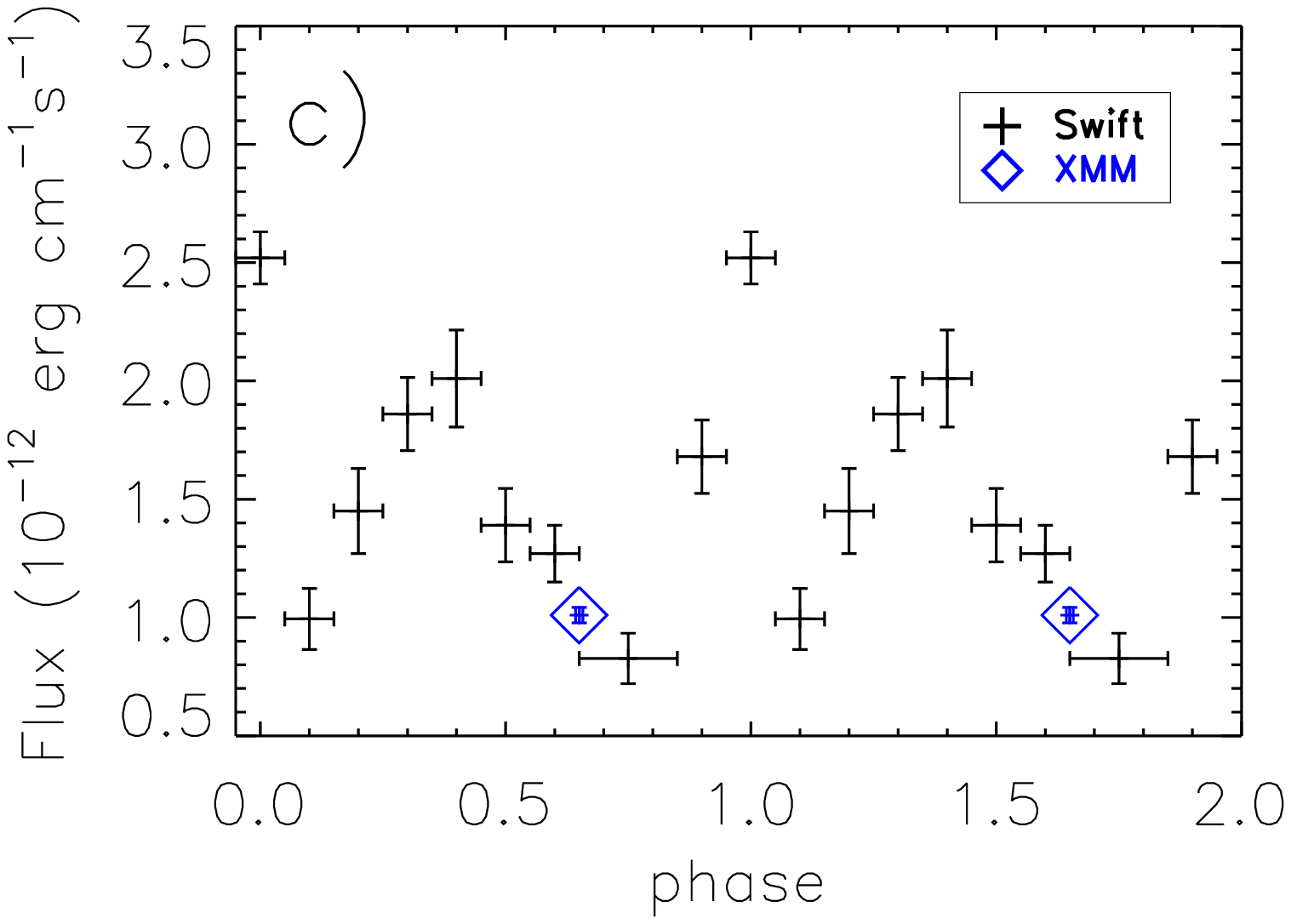} &
\hspace{-15.00mm}
\includegraphics[width=4. in]{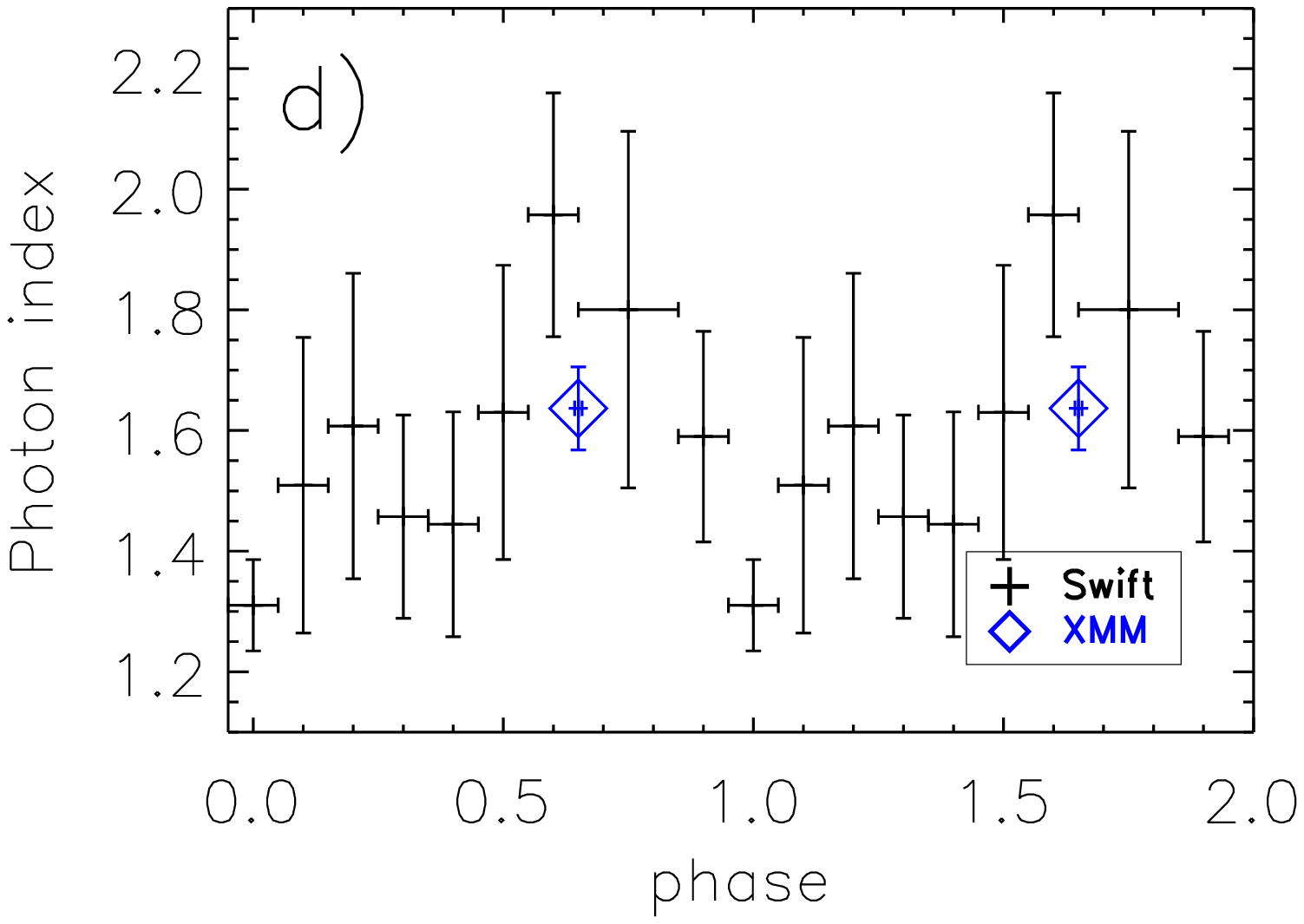} \\
\end{tabular}
\vspace{-6.00mm}
\figcaption{Observation times and count rates for {\em Swift} observations and results of the spectral analysis.
{\em a}) Unfolded {\em Swift} light curve in the 0.5--10 keV band. Vertical dashed lines indicate phase 0.
{\em b}) {\em Swift} 0.5--10 keV count rates versus orbital phase.
{\em c}) 0.5--10 keV absorption-corrected flux versus orbital phase.
{\em d}) Power-law photon index versus orbital phase.
\label{fig:spectra}
}
\end{figure*}

In order to measure the orbital period using the {\em Swift} data,
we employed epoch folding \citep{l87}, because it uses the count rates,
and thus takes care of the highly unequal
exposures of the observations. We folded the {\em Swift} light curves at different
test periods around the {\em Fermi}-measured value ($\pm 4$ days, step size=0.01 days) with the phase fixed
to zero at MJD 55403.3 \citep{fermi12}, and found the best period to be $P_{\rm orb}=16.57\pm0.11$ days
which is consistent with the {\em Fermi} measured value (see Fig.~\ref{fig:time}).
We also tried different binnings (4--10 bins) and different
step sizes (0.01--0.13 days) and found consistent results.
We note that the detection significance was marginal ($\sim 3\sigma$, see Fig.~\ref{fig:time}) even
with the known period (no search trials),
and thus it would have been very difficult to detect the X-ray modulation and measure the period 
without the guide of the gamma-ray measurement.

In Figures~\ref{fig:spectra}a and \ref{fig:spectra}b, we show the unfolded count rates as a function of time
in days since the {\em Fermi} epoch (MJD 55403.3), and the count rates folded in orbital phase for each
observation, respectively. We find that there is a sharp peak at phase $\sim$ 0 as reported by \citet{fermi12}.
To examine this in more detail, we select the brightest five fluxes and henceforth refer to them as ``flares.''
We tried to measure the time scales for the flares by using various temporal binnings for each flare observation.
We find no evidence that the flares occurred in a narrow time bin; rather each of the flares seems to be
longer than the observations (2--10 ks). {Note that large flares like those which occurred
in the first $\sim$ 100 days of observations were not observed in groups 2, 3 and 4 in spite of exposure at
phase 0 in all 3 groups (see Figs.~\ref{fig:spectra} a and b).}

We note that there are two low outliers at phase $\sim$ 0.3--0.4 in group 3
(blue in Fig.~\ref{fig:spectra}b). For the two, 11 and 13 events were collected in the
source region (see Section~\ref{spectrumana})
for 1.4-ks and 1.6-ks exposures, respectively. We checked if the photon collecting areas were
reduced for the two observations due to bad pixels in the source region, and found no significant reduction.
Therefore, we included them in the timing and spectral analyses.

\subsection{Spectral Analysis}
\label{spectrumana}
For the {\em Swift} data, we extracted the source spectra from a circle of radius 20$''$ and the backgrounds from
an annular region of inner radius 40$''$ and outer radius 80$''$ centered at the source position. For the observation
in which the source position was offset by $\sim9''$ compared with the {\em Chandra} position,
we shifted the source extraction region by such amount. The corresponding
ARFs were produced using {\ttfamily xrtmkarf} and corrected for the exposure using {\ttfamily xrtexpomap}.
Each spectrum had $\sim$~10--250 counts in it, and not all the individual spectra were useful for a meaningful
analysis. We folded the observations using 10 phase bins and the {\em Fermi} ephemeris because the latter
is more precise than that measured in this work (see Section~\ref{timingana}). We then combined the
observations in each phase bin for the spectral analysis. Even after combining spectra, there were not enough events
in some phases. Therefore we had to further combine orbital phase bins 7 and 8; see Table~\ref{ta:spec}.

For the {\em XMM-Newton} data, we extracted the source spectrum from circular regions having radius of
16$''$ and background spectra from source-free regions having radius of 32$''$ on the same chip.
Corresponding response files were produced using the {\ttfamily rmfgen} and the {\ttfamily arfgen} tasks
of SAS 11.0.0. The spectrum was then grouped to have a minimum of 20 counts per bin.

We used {\ttfamily XSPEC} 12.7.1 to fit the spectra. We first fit the {\em XMM-Newton} data
(MOS1, MOS2 and PN) with a simple absorbed
power law ({\ttfamily tbabs*pow}), an absorbed blackbody ({\ttfamily tbabs*bbody}) and an absorbed thermal
bremsstrahlung model ({\ttfamily tbabs*bremss}). The power-law and the bremsstrahlung models fit the spectrum
well ($\chi^2$/dof = 95.87/120, 97.00/120, respectively), and the residuals from the fit were featureless.
Although both the power-law and the bremsstrahlung models were acceptable, we report the power-law model, since it
gives a slightly better fit and is more commonly used for other similar binary systems.
The power-law fit parameters we obtained for the {\em XMM-Newton} data are consistent with the previously
reported values \citep{pmk+11, hess12}.

For the {\em Swift} data, we attempted to fit the spectra using the usual chi-squared statistics.
However, when there were insufficient events, we used the C-statistic implemented as {\ttfamily cstat} in
{\ttfamily XSPEC} %
without binning the spectrum (in the 0.5--10 keV band). We checked if the C-statistic fit results are comparable to the
$\chi^2$ results with two {\em Swift} observations that have enough counts (phases 0 and 3, see Table~\ref{ta:spec}),
and find that they agree within the statistical uncertainties.
We fit the data with an absorbed power-law model.

Due to the paucity of counts, we could not measure the hydrogen column density ($N_{\rm H}$) well for each orbital
phase. However, we find that the $N_{\rm H}$ values measured (i) with the archival {\em XMM-Newton}
observation (see Table~\ref{ta:spec}) and (ii) with the
{\em Swift} spectrum at phase 0, separated by $\sim$ 2 years from the {\em XMM-Newton} observation,
are consistent with each other, although
the uncertainty in the {\em Swift} value of $N_{\rm H}$ was relatively large.
Note that only 6--10\% of variation
in $N_{\rm H}$ has been seen in similar sources over a long period
\citep[e.g., LS~5039 and PSR~B1259$-$63;][]{tku+09,utt+09}. Furthermore, large
orbital variations in $N_{\rm H}$ have only been seen in accreting systems \citep[e.g.,][]{mcr09},
whereas 1FGL~J1018.6$-$5856 shows no evidence of accretion, e.g., shows no features in its spectrum.
Therefore, we fixed the value of $N_{\rm H}$ to that
measured with {\em XMM-Newton}. The fit results are summarized in Table~\ref{ta:spec} and plotted
in Figure~\ref{fig:spectra}d.

\begin{figure}
\centering
\hspace{-6mm}
\includegraphics[width=3.6 in]{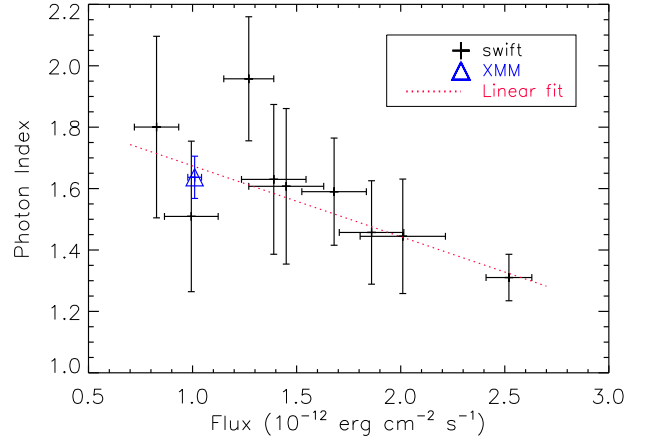} 
\hspace{-4mm}
\vspace{-3mm}
\figcaption{Correlation between 0.5--10 keV flux and photon index.
\label{fig:corr}
}
\hspace{1mm}
\vspace{-2mm}
\end{figure}

In Figures~\ref{fig:spectra}b and \ref{fig:spectra}c, we see the phase 0 flux flares
reported by \citep{fermi12}, but only in the first ~100 days of observations.
To understand the properties of the flares (the five brightest points), we fit the combined
spectrum of the flares (noted as ``flare'' in Fig.~\ref{fig:spectra}a and \ref{fig:spectra}b
and compare it with that of the remaining data in the same phase bin.
For the flares, the combined spectrum was well fitted with an absorbed power-law with photon index of 1.30(9),
and the flux was $3.12(16)\times 10^{-12}\ \rm erg\ cm^{-2}\ s^{-1}$.
For the rest of the data in phase 0, the photon index was 1.50(16), and the flux
was $1.49(13)\times 10^{-12}\ \rm erg\ cm^{-2}\ s^{-1}$.
The very hard spectrum in phase 0 seems to be driven by the five flares since
those have most of the events, and both the separated spectra (five flares and the rest)
in phase 0 seem to fit in the hardness/flux correlation trend individually (see Fig.~\ref{fig:corr} and below).
However, we note that the difference in the photon indexes is not statistically significant.

We find evidence of a negative correlation between the flux and the spectral hardness
(see Figs.~\ref{fig:spectra}c, \ref{fig:spectra}d and \ref{fig:corr}).
In order to quantify the significance of the putative correlation,
we calculated the Spearman's rank order
correlation coefficient. The rank order coefficient was
$r_{\rm s}=-0.77$ ($r_{\rm p}=-0.71$ for Pearson's product-moment correlation coefficient)
for 10 data samples, implying $\sim3.4\sigma$ ($\sim2.4\sigma$) significance for the (linear) correlation.
If we ignore phase 0 where the flaring activity dominates the persistent emission (i.e. in case the
flares are caused by a different physical process), the correlation coefficient
is $r_{\rm s}=-0.68$ ($r_{\rm p}=-0.63$),
implying $\sim2.5\sigma$ ($\sim 1.5\sigma$) significance.

We then conducted simulations to take into account the uncertainties in the flux
and photon index for the correlation. We first verified that the error contours for both
parameters were approximately elliptical Gaussian.
Since the parameters co-vary, we used the covariance matrices obtained during the spectral fits to
properly account for this effect in our simulations.
For each simulation, we varied the flux and photon index using Gaussian random numbers,
calculated the rank order correlation coefficient, and counted the occurrences of non-negative correlation.
The latter occurred 365 times in 10,000 simulations, suggesting the confidence level of the negative
correlation to be $\sim$ 96\% ($\sim$ 90\% if ignoring phase 0).

We also tried to fit the data to a constant function (e.g., no correlation)
or a linear function (e.g., negative correlation) taking into account both uncertainties in the
flux and the photon indices (Fig.~\ref{fig:corr}). The null hypothesis probability for the constant
function was $\sim$ 5\% ($\chi^2$/dof=16.9/9), and adding a linear slope improved the fit significantly
(F-test probability of 0.001). The linear fit was acceptable
with a null hypothesis probability 86\% ($\chi^2$/dof=3.9/8),
and we measured the slope to be $-0.23 \pm 0.07$ (per $10^{-12}\ \rm erg\ cm^{-2}\ s^{-1}$).
We therefore conclude that there is evidence of anti-correlation between the spectral index and the flux
but that additional data will be required to verify it.

\section{Discussion}
\label{sec:disc}
We find that the X-ray flux of 1FGL~J1018.6$-$5856 shows orbital modulation and
evidence of being correlated with spectral hardness.
We also find that the period of the X-ray orbital modulation is $16.57\pm0.11$ days, consistent
with that of the gamma-ray modulation. Further, we find that the average X-ray orbital light curve
is smoother than previously reported, but is punctuated by occasional high-flux ``flares'' near
orbital phase 0, and that the persistent peak of the orbital modulation in the X-ray flux
appears to be in the phase 0.3--0.4.

1FGL~J1018.6$-$5856 shares some X-ray properties with known gamma-ray binaries
LS~5039, LS~I~$+$61$^\circ$303 and PSR~B1259$-$63, where the compact star companion is either
known or generally assumed to be a neutron star.
The photon index of 1FGL~J1018.6$-$5856 in the 0.5--10 keV band varies between $\sim$ 1.3 and $\sim$ 2.0.
These values and this range are similar to those of the other sources.
Orbital variations of photon index for other gamma-ray binaries are 1.45--1.61 for LS~5039 \citep{tku+09},
1.35--1.83 for PSR~B1259$-$63 \citep{ktn+95, hck+99, utt+09}, and 1.7--2.0 for LS~I~$+$61$^\circ$303 \citep{ltz+11}.
Spectral variation with orbital phase is expected in models of gamma-ray binaries,
since any orbital eccentricity results in a varying separation between compact object and companion star,
along with a variable relative shock distance and particle/photon flux at the shock
location \citep[e.g.][]{tak94, ta97, d06, bkk+08}.
In such models, we naively expect the spectral variability to be more pronounced for a source with
large eccentricity. Indeed, the orbital variation of the
spectral photon index is stronger for larger eccentricity in case of the three sources, LS~5039,
LS~I~$+$61$^\circ$303, and PSR~B1259$-$63 in Table~\ref{ta:list}. However, 1FGL~J1018.6$-$5856 does
not follow this trend; its eccentricity has been argued to be small \citep{fermi12} but
the spectral variation is large, 
which is puzzling. However, the current measurements have large uncertainties so require verification
with more precise measurements before drawing final conclusions. We note that shock viewing geometry
could also play a role in variable flux and spectral parameters, even in a circular orbit.

\begin{figure}
\centering
\vspace{6mm}
\begin{tabular}{cc}
\hspace{-1.75mm}
\vspace{-4.65mm}
\includegraphics[width=3.117 in]{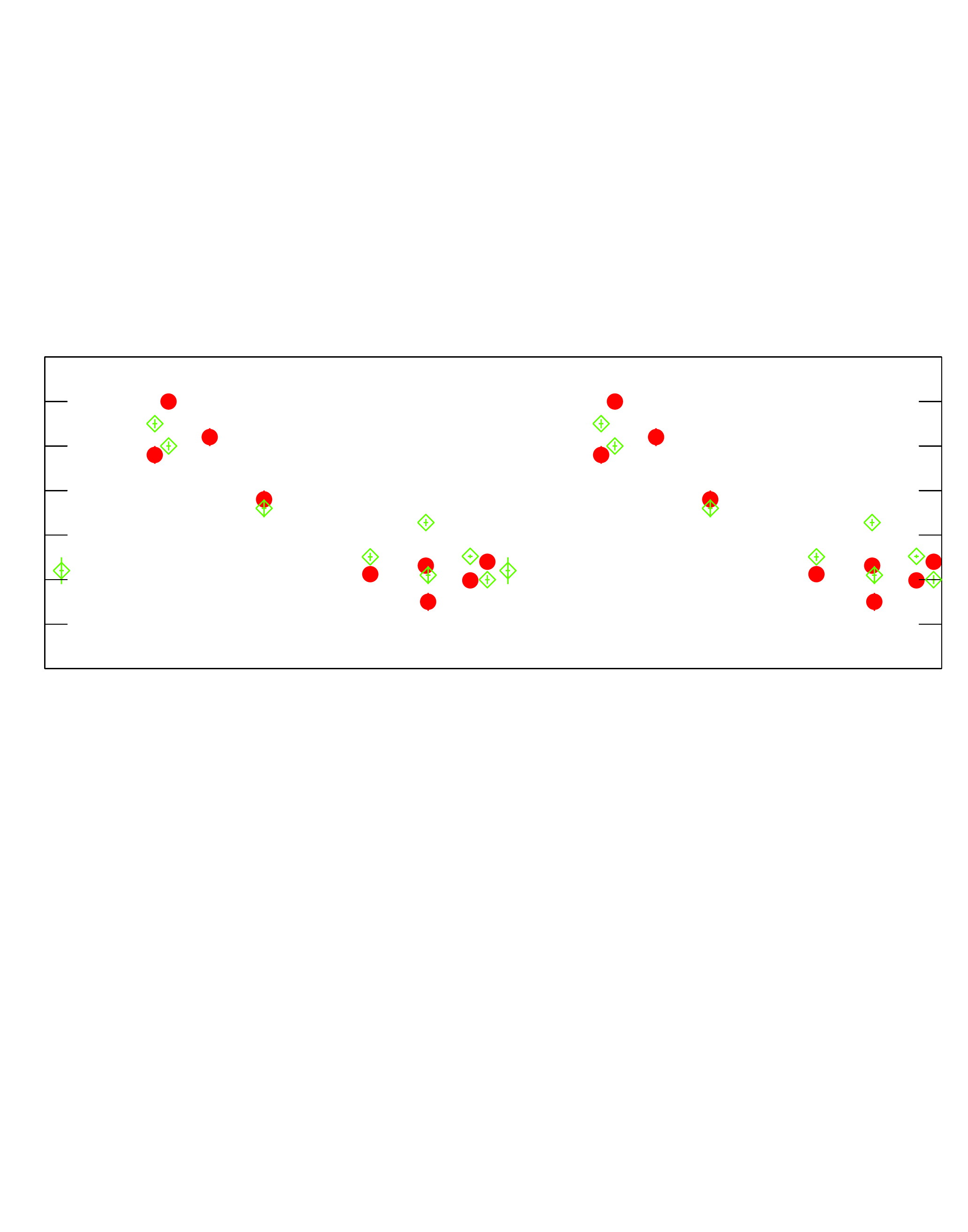} \\
\hspace{-10.09mm}
\vspace{-8.47mm}
\includegraphics[width=3.71 in]{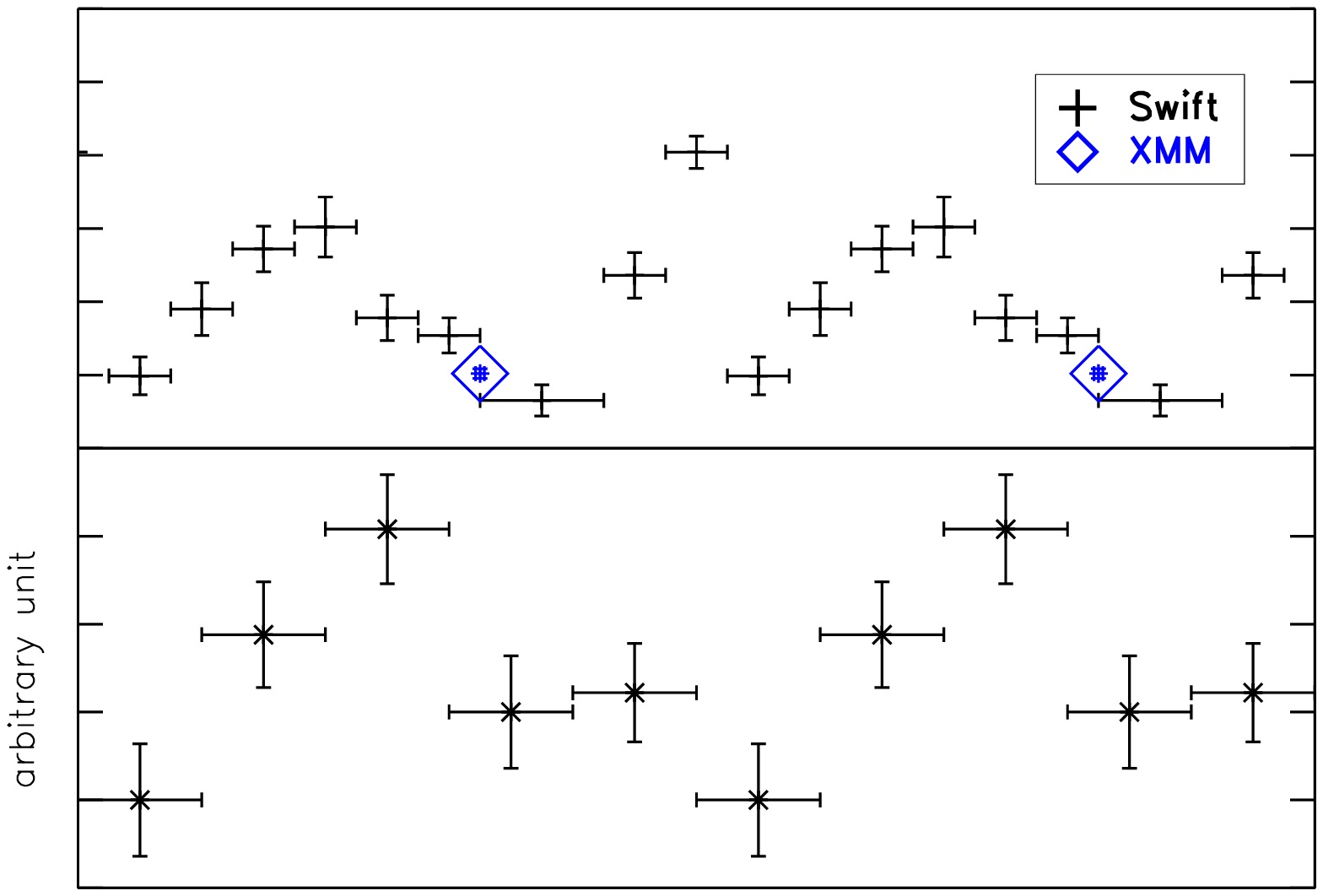} \\
\hspace{-0.9mm}
\vspace{1.00mm}
\includegraphics[width=3.18 in]{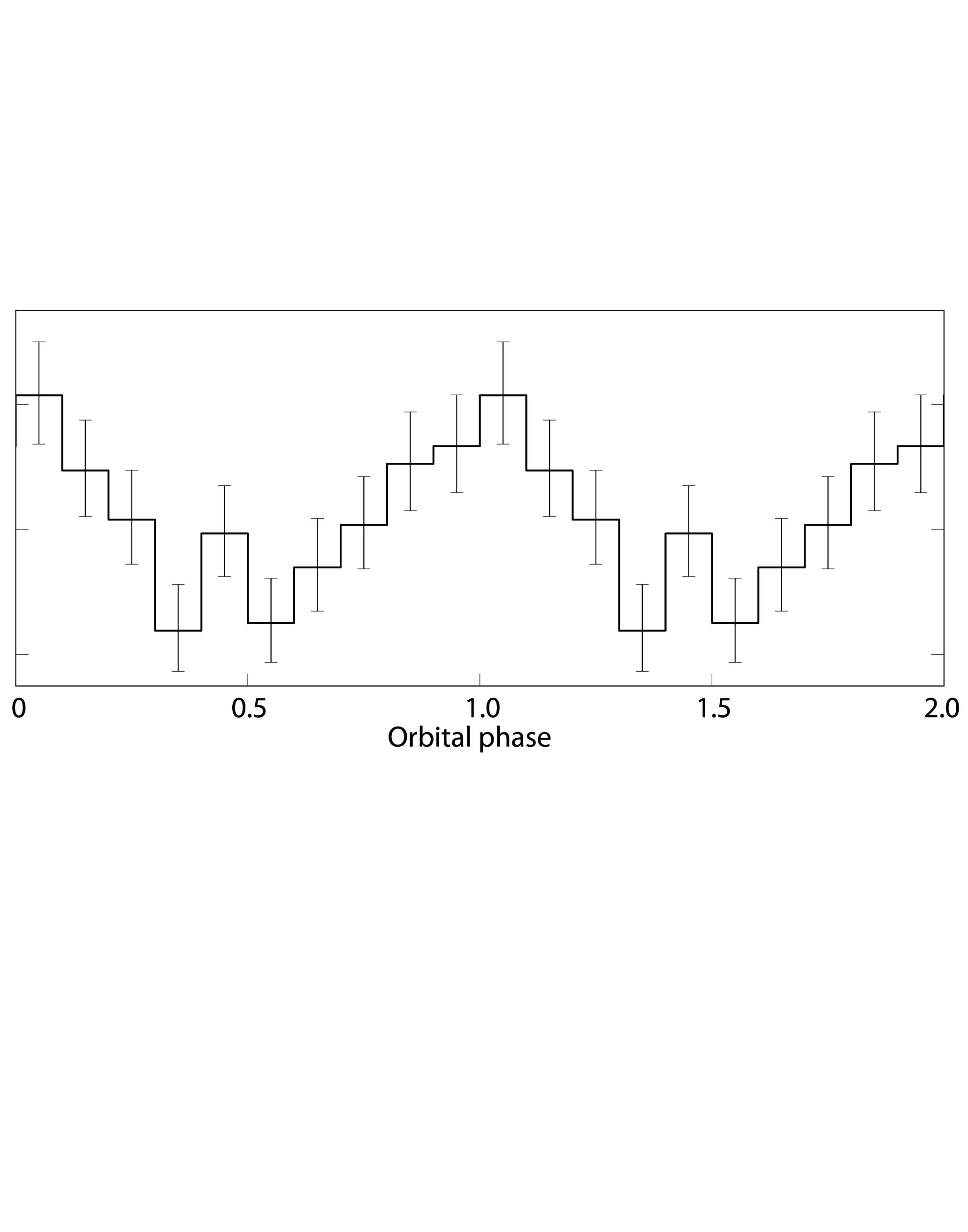} \\
\end{tabular}
\vspace{-3.0mm}
\figcaption{Orbital modulation of radio, X-ray, and Gamma-ray fluxes.
From top, radio data \citep[diamonds at 9 GHz and circles at 5.5 GHz, figure taken from][]{fermi12},
0.5--10 keV absorption-corrected flux (from this work),
count rates in the 18--40 keV band \citep{ltc+11}, and 1--10 GeV flux \citep[figure taken from][]{fermi12}.
\label{fig:radiotohardx}
}
\end{figure}

We find that the spectral index of 1FGL~J1018.6$-$5856 shows evidence of being
anti-correlated with the 0.5--10 keV
flux (see Fig.~\ref{fig:corr}). The same trend was also observed in two of the gamma-ray binaries above
\citep[LS~I~$+$61$^\circ$303, LS~5039;][]{ltz+11, tku+09} but not in
PSR~B1259$-$63 \citep[][]{ktn+95, hck+99, utt+09}.
Why such a correlation should be present in some systems, but not all, if indeed they all have a common
nature, is puzzling.

\citet{tak94} and \citet{ta97} proposed a pulsar wind/stellar wind interaction model for PSR~B1259$-$63.
In the model, the locations of the termination shock as a function of orbital phases are determined
by orbital geometry and pressure balance of the two winds. The time scales of various physical
processes (e.g., particle acceleration, the synchrotron radiation and the inverse Compton processes)
and spectral parameters are then calculated.
\citet{ta97} noted that different interaction models can be best tested using the time behavior of the
X-ray luminosity and spectrum, and they demonstrated the X-ray flux and spectral variability with changing
orbital phase for PSR~B1259$-$63.

For 1FGL~J1018.6$-$5856, the stellar wind outflow \citep[for O6V star companion;][]{pkh+96} may be
larger than but different in geometry from that in PSR~B1259$-$63 (which has a Be star companion), where
the mass outflow rate is smaller but is more concentrated in the equatorial plane for the case of
a Be star \citep[][]{bc93}. Considering this, the effective mass outflow
parameter \citep[$\Upsilon/f$,][]{tak94} for an O6V star can be comparable to or larger
than that of a Be star. In such a case, the model predicts that the
dominant physical process would be synchrotron radiation, consistent with what we infer from
the correlation between radio and X-ray flux below. Although it may be possible to apply the model
to 1FGL~J1018.6$-$5856 given a detailed orbital geometry, the latter is not yet constrained
so detailed modeling cannot presently be done.

We note that our observed hardness/flux correlation cannot be due to orbital variation of $N_{\rm H}$.
If it were, one would expect the count rate to be smaller for a harder
spectrum, which we do not observe.
If we assume that $N_{\rm H}$ varies by as much as 10\% for an orbit as has been seen in other
gamma-ray binaries \citep[e.g., LS~5039 and PSR~B1259$-$63;][]{tku+09,utt+09},
the variation can affect the photon index by $\sim$ 5\%, which does not explain the much larger
observed spectral variation.

\citet{m12} categorized gamma-ray binaries into three types based on radiative behavior and nature of the companion.
According to the categorization, 1FGL~J1018.6$-$5856 should be similar to LS~5039.
\citet{fermi12} argue that the two systems may differ based on the hardness/flux correlation in the gamma-ray band
and the fact that the X-ray maximum coincides with the gamma-ray maximum for 1FGL~J1018.6$-$5856 but not for LS~5039. 
However, we show that the large X-ray peak at phase 0 reported by \citet{fermi12} is likely to be caused
by occasional flaring behavior and is not obviously a persistent feature.
The maximum of the sinusoidal modulation in the X-ray band lies at phase 0.3--0.4 which does not coincide
with the 1--10 GeV gamma-ray peak (see Fig.~\ref{fig:radiotohardx}).
Thus, the orbital phase offset between the X-ray ($<$10 keV) band and gamma-ray band (1--10 GeV) is common to both
1FGL~J1018.6$-$5856 and LS~5039.

We note that although the very bright ``flares'' previously reported near phase 0 for this source,
in which the soft X-ray flux was seen to increase by factors of 3--5, do not appear to be a persistent
feature, some flux enhancement at that orbital phase is often present (see Figs.~\ref{fig:spectra}a and b).
Indeed, at phase 0, the X-ray flux is above the sinusoidal trend most of the time.
However, no significant flux increase in the radio or hard X-ray band at this orbital phase
has been observed \citep{fermi12, ltc+11}.
This is understandable if the flare amplitude during the radio and hard X-ray observations was small,
and/or the observations did not sample the ``narrow'' flare phase well enough to make a sensitive detection of the flare.
Nevertheless, the sinusoidal phase
of flux modulations in the radio, soft X-ray ($<$ 10 keV) and hard X-ray (18--40 keV) bands are
relatively well aligned (see Fig.~\ref{fig:radiotohardx}),
implying that they are all misaligned with the gamma-ray phase. The phase alignment between the radio
and the X-ray band may imply that the X-ray emission mechanism is the synchrotron process unlike
in LS~I$+$61$^\circ$303, where an offset between the radio and X-ray phases was observed,
and the X-ray emission was suggested to be due to an inverse Compton process \citep{hrl+00}.

Similar X-ray flares have been also seen in other systems
\citep[e.g., LS~I$+$61$^\circ$303 and LS~5039,][]{ltz+11,ktu+09}.
The flares in LS~I$+$61$^\circ$303 are aperiodic with kilo-second time scale \citep{ltz+11}, and
those in LS~5039 seem to be periodic with 10--20 ks time scale \citep[][]{ktu+09}. Although we were not able
to clearly characterize the time scale of the flares in 1FGL~J1018.6$-$5856, they seem to be periodic, and
the duration is rather long ($\gapp$ 2--10 ks), similar to those of LS~5039 but with relatively larger amplitudes.

The two systems, 1FGL~J1018.6$-$5856 and LS~5039, share many properties such as a flux/hardness correlation in
the soft X-ray band, phase alignment between the soft
and the hard X-ray band \citep{ltc+11}, and misalignment between the X-ray and GeV gamma-ray orbital phase.
However, they show different flux/hardness correlations in the gamma-ray band.
Detailed modeling and broadband observations in the future will help us to clearly tell
whether or not the two systems are different in nature.

\section{Conclusions}
\label{sec:concl}
We have analyzed {\em Swift}, {\em XMM-Newton} and {\em Chandra} data for the gamma-ray binary 1FGL~J1018.6$-$5856,
and find the orbital period of the X-ray ($<$10 keV) flux to be 16.57$\pm$0.11 days,
consistent with the value measured in the gamma-ray band.
We also show that the previously reported very large flux increase at phase 0 (factors of $\sim$ 3--5) occurred
only for the first $\sim$ 100 days of the {\em Swift} observations although substantial increases in flux
(a factor of $\leq 2$) are seen frequently at that phase. The persistent maximum of the X-ray orbital
modulation seems to occur at phase 0.3--0.4 and is significantly misaligned with the 1--10 GeV gamma-ray peak.
Finally, we show evidence that 1FGL~J1018.6$-$5856 exhibits a correlation between spectral hardness and the flux in the
0.5--10 keV band, which is common to several gamma-ray binaries and can hopefully be used to help 
understand the nature of X-ray emission from these interesting objects. \\

This research has made use of data obtained from the High Energy Astrophysics Science Archive Research Center
(HEASARC), provided by NASA's Goddard Space Flight Center.
V.M.K. acknowledges support
from an NSERC Discovery Grant, the FQRNT Centre de Recherche Astrophysique du Qu\'ebec,
an R. Howard Webster Foundation Fellowship from the Canadian Institute for Advanced
Research (CIFAR), the Canada Research Chairs Program and the Lorne Trottier Chair
in Astrophysics and Cosmology.

\end{document}